\documentclass[logo,11pt,a4paper]{ETHpaper}
\usepackage{graphicx, amsmath, amssymb,color,wasysym}
\usepackage{bm}
\usepackage[square,numbers,sort&compress]{natbib}
\usepackage{hyperref}
\usepackage[subrefformat=parens,labelformat=parens]{subcaption}

\usepackage{tikz}
\usetikzlibrary{calc}

\usepackage{longtable}

\begin{document}

\newcommand{\mean}[1]{\left\langle #1 \right\rangle}
\newcommand{\abs}[1]{\left| #1 \right|}
\newcommand{\ul}[1]{\underline{#1}}
\renewcommand{\epsilon}{\varepsilon}
\newcommand{\eps}{\varepsilon}
\renewcommand*{\=}{{\kern0.1em=\kern0.1em}}
\renewcommand*{\-}{{\kern0.1em-\kern0.1em}}
\newcommand*{\+}{{\kern0.1em+\kern0.1em}}

\newcommand{\potentiality}{H^{\mathrm{norm}}}
\newcommand{\RA}{\Rightarrow}
\newcommand{\bbox}[1]{\mbox{\boldmath $#1$}}
\newcommand{\code}[1]{\texttt{#1}}
\newcommand{\TODO}[1]{\textcolor{red}{#1}}
\newcommand{\N}{\mathbb N}

\title{What is the Entropy of a Social Organization?}

\titlealternative{What is the Entropy of a Social Organization?}

\renewcommand*{\thefootnote}{\fnsymbol{footnote}}
\author{Christian Zingg\footnote{czingg@ethz.ch}, Giona Casiraghi\footnote{gcasiraghi@ethz.ch}, Giacomo Vaccario\footnote{gvaccario@ethz.ch}, Frank Schweitzer\footnote{fschweitzer@ethz.ch}}
\authoralternative{C. Zingg, G. Casiraghi, G. Vaccario, F. Schweitzer}
\address{Chair of Systems Design, ETH Zurich, Weinbergstrasse 58, 8092 Zurich, Switzerland}

\reference{(Submitted for publication)} 
\www{\url{http://www.sg.ethz.ch}}

\makeframing
\maketitle
\renewcommand*{\thefootnote}{\arabic{footnote}}\setcounter{footnote}{0}

\begin{abstract}
	We quantify a social organization's potentiality, that is its ability to attain different configurations.
	The organization is represented as a network in which nodes correspond to individuals and  (multi-)edges to their multiple interactions.
        Attainable configurations are treated as realizations from a network ensemble.
To have the ability to encode interaction preferences, we choose the
        generalized hypergeometric ensemble of random graphs,
        which is described by a closed-form probability distribution.
        From this distribution we calculate Shannon entropy as a measure of potentiality.
        This allows us to compare different organizations as well different stages in the development of a given organization.
        The feasibility of the approach is demonstrated using data from 3 empirical and 2 synthetic systems.
  \end{abstract}

\section{Introduction}\label{sec:introduction}

Social organizations are ubiquitous in our everyday life, ranging from project teams, e.g. to produce open source software \citep{Schweitzer2014} to special interest groups, such as sports clubs~\citep{Zachary1977} or conference audiences~\citep{Isella2011} discussed later in this paper.
Our experience tells us that social organizations are highly dynamic.
Individuals continuously enter and exit, and their interactions change over time. Characteristics like these make social organizations complex and difficult to quantify.

Network science allows to study such complex systems in terms of networks, where nodes represent individuals and edges their interactions~\cite{Dorogovtsev2003,Newman2003}.
Under this assumption, a social organization can be represented by a \textbf{network ensemble}.
Every network in this ensemble corresponds to one possible configuration of interactions in this organization.
Thus, the network that we can reconstruct from observed interactions is only one particular realization from this ensemble.
Other configurations can also be realized  with a given probability.
To cope with this, we need a probability distribution that characterizes the network ensemble and reflects its constraints, such as given numbers of nodes or interactions, or preferences for interactions.
The probability space defined by this distribution can then be seen as a way to quantify the \emph{number} and the \emph{diversity} of possible states of a social organization.
We argue that such possible states give an indication of the \emph{potentiality} of this organization, i.e., its ability to attain different configurations.

This, however, is an unsolved problem.
First, we need to decide about a probability distribution suitable for reflecting the interactions and constraints in social organizations.
Second, based on this distribution we need to quantify the diversity of the organization.
To solve the first task, in this paper we utilize the hypergeometric ensemble, as explained in Sect.~\ref{sec:ensembleFitting}.
To solve the second task, we compute the \textbf{information entropy} of this ensemble, as shown in Sect.~\ref{sec:tractabilityOfTheEntropy}.

Information entropy~\citep{Ebeling1998} has recently gained popularity in the study of social organizations.
Shannon entropy has, for example, been applied to study communication networks~\citep{Zhao2011}, human contact events~\citep{Kulisiewicz2018}, or anomalies in computer networks on a campus~\citep{Santiago-Paz2012}.
By generalizing the concept of entropy, even complexity measures~\citep{Rajaram2016} or classifications for complex systems~\citep{Hanel2011} have been suggested.
Finally, entropies have also been applied in combination with network ensembles to analyze complexity behind structural constraints~\citep{Bianconi2009} or spatial constraints~\citep{Coon2018} or how restrictive topological constraints are~\citep{Bianconi2008}.

Recent works that combine network ensembles and entropy analyze the effect of \textit{predefined} constraints.
For example, in \citep{Bianconi2008}, the author studies how entropy changes when fixing degree sequences or community structures, that are derived from the network topology.
By enforcing such topological constraints, the resulting ensembles serve as null models to study the expected level of order given the fixed constraints. 
However, real systems are affected by a very large number of constraints and because they are so many, they cannot be easily predefined. 
To consider also these constraints, we use the generalized hypergeometric ensemble, gHypEG \citep{Casiraghi2018}.
This allows to encode observed interaction preferences among every pair of individuals as biases in the edge formation.
By this, we capture in the network ensemble \textit{any} constraints that manifest as interaction preferences between individuals.

The paper is organized as follows.
In Sect. \ref{sec:potentiality} we derive our measure of potentiality for a social organization based on the Shannon entropy of a probability distribution.
In Sect. \ref{sec:ensembles} we first explain how this probability distribution can be calculated for a Generalized Hypergeometric Ensemble.
We then also show how to obtain the Shannon entropy of this distribution by means of a limit-case approximation, because direct computations are infeasible because of the large size of the ensemble.
In Sect. \ref{sec:applications} we measure the potentiality of \(3\) empirical social organizations and then compare the computed values across the organizations.
Finally, in Sect. \ref{sec:conclusion} we summarize our approach and comment on its strengths and weaknesses.

\section{Quantifying the Potentiality of a Social Organization}\label{sec:potentiality}

\subsection{Network Representation of a Social Organization}
\label{sec:netw-repr-soci}

We adopt a network perspective to study social organizations.
The nodes of the network represent individuals, and the observed interactions between them are represented by edges. If multiple interactions between the same pair of individuals occur, we consider them as separate edges, so-called \emph{multi-edges}~\cite{Bollobas1998}.
For simplicity, we will always refer to them as edges.
In this article, we will focus on undirected edges without self-loops, however, the methodology discussed can easily be extended to encompass directed edges, and self-loops.

According to this perspective, the observation of a social organization composed of $n$ individuals yields an  network $\hat g$, with $n$ nodes and $m$ edges, where $m$ is the number of observed interactions.
The \emph{state} of the social organization, instead, is defined by a network ensemble composed of all possible networks $S = \{g_0,\,\dots,\,g_N\}$, that encompass all possible configurations the social organization could attain, with $\hat g\in S$.

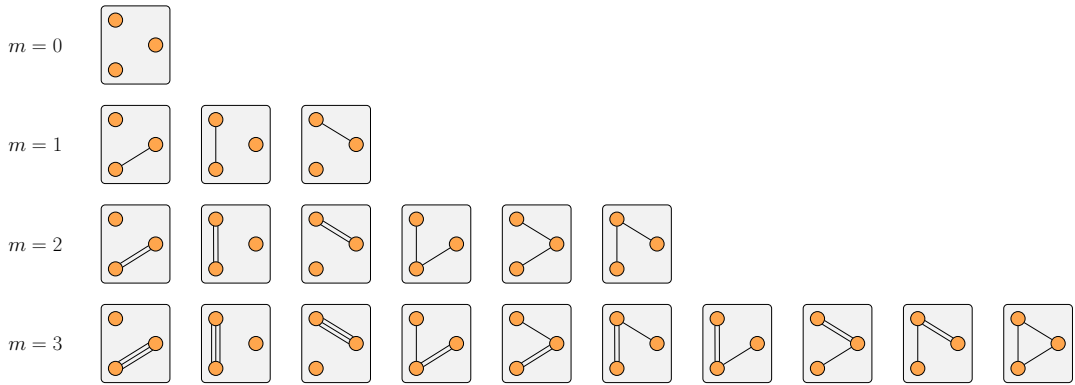
\begin{figure}[htbp]
	\begin{center}
		\resizebox{!}{0.25\textheight}{

\newcommand{\arrowbundle}[3]{
	\ifnum#3>0  
		\def\increment{35}  
		\pgfmathtruncatemacro\iseven{{Mod(#3,2)}}  
		\ifnum\iseven=0
			\pgfmathsetmacro\baseangle{(\increment / 2) + (((#3 / 2) - 1) * \increment)}
		\else
			\pgfmathsetmacro\baseangle{(#3 - 1) / 2 * \increment}
		\fi

		\pgfmathanglebetweenpoints{\pgfpointanchor{#1}{center}}{\pgfpointanchor{#2}{center}}
		\pgfmathsetmacro\abangle{ \pgfmathresult }  
		\pgfmathanglebetweenpoints{\pgfpointanchor{#2}{center}}{\pgfpointanchor{#1}{center}}
		\pgfmathsetmacro\baangle{ \pgfmathresult }  

		\pgfmathtruncatemacro\lastarrowindex{#3 - 1}
		\foreach \x in {0,...,\lastarrowindex} {
			\pgfmathsetmacro\currangleab{ \abangle + \baseangle - \x * \increment }
			\pgfmathsetmacro\currangleba{ \baangle + \baseangle - (\lastarrowindex - \x) * \increment }
			\draw[arrow] (#1.\currangleab) -- (#2.\currangleba);
		}
	\fi
}

\newcommand{\statespaceboundingbox}[2]{
	\node[rectangle, draw, fill=black!5, inner sep=5, rounded corners, #1] {#2};
}

\newcommand{\statespacenetwork}[3]{
	\begin{tikzpicture}
		\tikzstyle{vertex}=[circle, draw, inner sep=5, fill=orange!70]
		\tikzstyle{arrow}=[thick]

		\node[vertex] (vlower) {};
		\node[vertex, yshift=50] (vupper) at (vlower) {};
		\node[vertex, xshift=40] (vmid) at ($(vlower)!0.5!(vupper)$) {};

		\arrowbundle{vlower}{vmid}{#1}
		\arrowbundle{vlower}{vupper}{#2}
		\arrowbundle{vmid}{vupper}{#3}
	\end{tikzpicture}
}


\begin{tikzpicture}
	\tikzstyle{boundingbox}=[rectangle, draw, fill=black!5, inner sep=7, rounded corners]
	\node[boundingbox] (network0) {\statespacenetwork{0}{0}{0}};

	\node[boundingbox, yshift=-100] (network1a) at (network0) {\statespacenetwork{1}{0}{0}};
	\node[boundingbox, xshift=100] (network1b) at (network1a) {\statespacenetwork{0}{1}{0}};
	\node[boundingbox, xshift=100] (network1c) at (network1b) {\statespacenetwork{0}{0}{1}};

	\node[boundingbox, yshift=-100] (network2a) at (network1a) {\statespacenetwork{2}{0}{0}};
	\node[boundingbox, xshift=100] (network2b) at (network2a) {\statespacenetwork{0}{2}{0}};
	\node[boundingbox, xshift=100] (network2c) at (network2b) {\statespacenetwork{0}{0}{2}};
	\node[boundingbox, xshift=100] (network2d) at (network2c) {\statespacenetwork{1}{1}{0}};
	\node[boundingbox, xshift=100] (network2e) at (network2d) {\statespacenetwork{1}{0}{1}};
	\node[boundingbox, xshift=100] (network2f) at (network2e) {\statespacenetwork{0}{1}{1}};

	\node[boundingbox, yshift=-100] (network3a) at (network2a) {\statespacenetwork{3}{0}{0}};
	\node[boundingbox, xshift=100] (network3b) at (network3a) {\statespacenetwork{0}{3}{0}};
	\node[boundingbox, xshift=100] (network3c) at (network3b) {\statespacenetwork{0}{0}{3}};
	\node[boundingbox, xshift=100] (network3d) at (network3c) {\statespacenetwork{2}{1}{0}};
	\node[boundingbox, xshift=100] (network3e) at (network3d) {\statespacenetwork{2}{0}{1}};
	\node[boundingbox, xshift=100] (network3f) at (network3e) {\statespacenetwork{0}{2}{1}};
	\node[boundingbox, xshift=100] (network3g) at (network3f) {\statespacenetwork{1}{2}{0}};
	\node[boundingbox, xshift=100] (network3h) at (network3g) {\statespacenetwork{1}{0}{2}};
	\node[boundingbox, xshift=100] (network3i) at (network3h) {\statespacenetwork{0}{1}{2}};
	\node[boundingbox, xshift=100] (network3j) at (network3i) {\statespacenetwork{1}{1}{1}};

	\node[xshift=-100] (label0) at (network0) {\huge\(m=0\)};
	\node at (label0 |- network1a) {\huge\(m=1\)};
	\node at (label0 |- network2a) {\huge\(m=2\)};
	\node at (label0 |- network3a) {\huge\(m=3\)};
\end{tikzpicture}
		}
		\end{center}
	\caption{
		Visualization of the possible networks for \(3\) nodes and different numbers of edges.
		The edges are undirected and self-loops are not considered.
		}\label{fig:ensembleIllustration}
\end{figure}

As an example, Fig. \ref{fig:ensembleIllustration} illustrates every possible network for \(3\)~nodes and an increasing number of edges $m$.
While for \(3\)~nodes and \(2\)~edges there are \(6\) possible networks, for 3 edges already 10 networks result.
For \(10\)~nodes and \(10\)~edges there would be more than \(2\cdot10^{10}\)~possible networks.
The general expression for the number of possible networks is
\begin{equation}\label{eq:networkSpaceSize}
	{\frac{n(n-1)}{2}+m-1\choose m}
\end{equation}
where \(n(n-1)/2\) denotes the number of combinations between $n$ nodes.
Equation~\eqref{eq:networkSpaceSize} can be derived directly from the known formula for drawing unordered samples with replacement.
The replacement is important because we consider multi-edges.

Notwithstanding the large number of possible networks, not all of them appear with the same probability.
We denote with $g$ a particular network, and by $P(g)$ the probability to find $g$, given $n$ and $m$.
A proper expression for $P(g)$ has to take into account that the ensemble, in addition to a fixed number of nodes and edges, also may have other constraints that need to be reflected in the probability distribution.
This issue will be further discussed in Sect.~\ref{sec:ensembles}.
But, assuming that we have such an expression for $P(g)$, the information from this can be compressed by calculating Shannon entropy~\citep{Shannon1948a}
\begin{equation}\label{eq:entropy}
	H = -\sum_{g\in S} P(g)\log{P(g)}
\end{equation}
where \(S\) denotes the set of all possible networks for fixed $n$, $m$.

\subsection{Potentiality of a Social Organization}
\label{sec:potent-soci-organ}
\paragraph{Potentiality and constraints.}

In our network representation, a large number of possible networks translates into a large number of possible configurations that can be attained by the social organization.
Hence, we can use entropy to characterize the \emph{potentiality} of the social organization, that is, its ability to attain these \emph{different configurations} under the existing constraints.
These constraints limit the number of configurations, i.e they reflect that a social organization cannot change from a given configuration to any arbitrary other configuration.
Thus, \emph{constraints lower the potentiality} of social organizations.

Such constraints can be temporal, i.e., they impose an order of occurrence to the edges in the network, as extensively examined in~\citep{Scholtes2017,Kulisiewicz2018}.
Or there can be spatial constraints that restrict the individuals in the choice of communication partners
~\cite{Vaccario2018,Liben-Nowell2005}.
Social organizations can also be subject to hierarchical constraints~\cite{Zanetti2013}, restricting e.g. the flow of information, or to social constraints~\cite{Scholtes2016} as discussed in Sect. \ref{sec:applications}.

\paragraph{How to proxy constraints.}
We consider distributions $P(g)$ that capture communication biases among the individuals.
These biases, or preferences, are the consequences of the constraints that restrict the social organization.
We take the observed number of interactions between each pair of individuals in a defined time interval as the proxy for the constraints.
For this reason, we set the expected number of interactions between each pair of nodes in the ensemble to the observed ones.
This choice ensures that the distribution $P(g)$ encodes the constraints in the ensemble, because we assume that constraints are expressed in the number of interactions between the nodes.

In the next Section we will demonstrate how to specify the probability distribution \(P(g)\) characterizing the network ensemble such that this is achieved.
To do so, we will employ the generalized hypergeometric ensemble (gHypEG) developed by~\citet{Casiraghi2018}.

\paragraph{Network ensembles and their probability distribution.}
What have we obtained by calculating Shannon entropy, i.e. a \emph{single number} to characterize $P(g)$?
To fully understand this, we have to recapture what information the probability distribution actually contains.
$P(g)$ in fact characterizes the \emph{diversity} of potential networks, i.e. the possible network configurations that can appear under the given constraints encoded in $P(g)$.
We denote the totality of these configurations as the \emph{network ensemble}.
If there are only a \emph{few} network configurations possible, the ensemble is comparably \emph{small} and the resulting entropy is \emph{low}.
On the other hand, if many network configurations are possible, the ensemble becomes very large and the entropy is high.

\section{Introducing the Generalized Hypergeometric Ensembles}\label{sec:ensembles}

\subsection{Obtaining \(P(g)\)}\label{sec:ensembleFitting}

For the calculation of Shannon entropy, Eq.~\eqref{eq:entropy}, we implicitly assumed that \(P(g)\) is known.
There are mainly two candidates for $P(g)$ that fit our requirements.
One is the family of \emph{exponential random graphs}, also known as ERGMs~\citep{Krivitsky2017,Park2004}.
ERGMs follow an exponential distribution, thus it is possible to compute their Shannon entropy.
Moreover they can incorporate a broad set of properties and constraints~\citep{Morris2008} which can fit virtually any characteristics of observed networks.
However, ERGM fitting algorithms, especially when fitted to multi-edge networks, have a major drawback: they tend to not converge
and thus cannot be efficiently computed for large networks.

Additionally, they are only suited to consider a limited predefined set of constraints and are not flexible enough to consider other types.
Therefore, our choice is the second candidate, which is the \emph{generalized hypergeometric ensemble} of random graphs~\citep{Casiraghi2018,Casiraghi2016} (gHypEG).
This ensemble extends the configuration model (CM)~\citep{Molloy1995} by encoding complex topological patterns, while at the same time preserving expected degree sequences.

Specifically, gHypEG keeps the number of nodes and edges fixed.
However, different from the CM, the probability to connect two nodes depends not only on their (out- and in-) degrees (i.e., number of stubs), but also on an independent \emph{propensity} of the two nodes to be connected, which captures \emph{non-degree related effects} as explained in the following.

\paragraph{Parameters of a gHypEG.}
The distribution of networks in a gHypEG is formulated in terms of two sets of parameters.
The first set of parameters is represented in terms of the combinatorial matrix~\(\mathbf\Xi\) that encodes the CM\@.
That means the entries $\Xi_{ij}$ reflect all ways in which nodes $i$ and $j$ can be linked.
As will be explained later in an undirected network without self-loops this number is \(2\tilde{d}_i\tilde{d}_j\) for rescaled degrees \(\tilde{d}_i\), \(\tilde{d}_j\) of nodes \(i\), \(j\).

The second set of parameters is represented in terms of the propensity matrix~\(\mathbf\Omega\) which encodes preferences of nodes to be connected.
That means, propensities allow to constrain the configuration model such that given edges are more likely than others, independently of the degrees of the respective nodes.
This creates a bias which is expressed by the ratio between any two elements \(\Omega_{ij}\) and \(\Omega_{kl}\), i.e., the odds-ratio of observing an edge between nodes \(i\) and \(j\) instead of between \(k\) and \(l\).

The matrices \(\mathbf\Xi\) and \(\mathbf\Omega\) both have dimension~\(n\times n\), where \(n\) is the number of nodes.
The probability distribution that reflects the biased edge allocation described above is given by the multivariate Wallenius non-central hypergeometric distribution~\citep{wallenius1963}.
I.e.,  the probability of a network \(g\) in the gHypEG with parameters \(\mathbf\Xi\) and \(\mathbf\Omega\) is given as follows:
\begin{equation}
\label{eq:walleniusNet}
	P(g\lvert\mathbf\Xi,\mathbf\Omega)=\left[\prod_{i,j\in V, i< j}{\dbinom{\Xi_{ij}}{A_{ij}}}\right]
         \int_{0}^{1}{\prod_{i,j\in V, i<j}{\left(1-z^{\frac{\Omega_{ij}}{S_{\mathbf{\Omega}} }}\right)^{A_{ij}}}dz}
\end{equation}
with
\begin{equation}
\label{eq:walleniusSOmega}
	S_{\mathbf{\Omega}}= \sum_{i,j\in V,i<j} \Omega_{ij}(\Xi_{ij}-A_{ij}).
\end{equation}
Equation \eqref{eq:walleniusNet} and Eq. \eqref{eq:walleniusSOmega} hold for undirected networks without self-loops  ($i<j$).

\paragraph{Calculating \(\Xi{}\) for networks.}
We obtain the \(\Xi{}\) matrix for a given network according to Definition~4 and Lemma~3 in~\citep{Casiraghi2018}.
But, since in our applications there are no self-loops, we implement additional correction factors to preserve the expected degrees in the ensemble.
Specifically, we ensure that the expected degrees are equal to the degrees in the initial network.
The details can be found in App. \ref{sec:fittingXi}.
Our \(\Xi_{ij}\) are therefore
\begin{equation}\label{eq:XiComputation}
	\Xi_{ij} :=
		\begin{cases}
			2 (d_i \theta_i) (d_j \theta_j) & \text{if } i < j \\
			0 & \text{else}\\
		\end{cases}
\end{equation}
where \(d_i\), $d_{j}$ denote the degree of nodes~\(i\) and $j$ and the \(\theta_i\), $\theta_{j}$ denote the correction factors that ensure the expected degrees are preserved.
In this definition the diagonal elements are \(0\) because we do not allow for self-loops.
Also the entries in the lower triangular part are \(0\) to account for the networks being undirected.

\paragraph{Calculating \(\Omega{}\) for networks.}
We obtain the respective \(\Omega{}\) matrix \emph{for a given \(\Xi\) matrix} according to Corollary~7.3 in \citep{Casiraghi2018}.
Thereby we ensure that, in addition to the expected degrees, even the expected numbers of edges between \emph{all} pairs of nodes in the ensemble are equal to the respective numbers of edges in the initial network.
Hence, our \(\Omega{}_{ij}\) are
\begin{equation}\label{eq:OmegaComputation}
	\Omega_{ij} :=
		\begin{cases}
                  \frac{\displaystyle 1}{\displaystyle c}\log{\left(1-\frac{\displaystyle A_{ij}}{\displaystyle\Xi_{ij}}\right)} &
                  \text{if } i < j \\
			0 & \mathrm{else}
		\end{cases}
\end{equation}
where $A_{ij}$ is the number of edges between nodes $i$ and $j$, and $c$ is a multiplicative constant which we choose such that the values in $\Omega$ are between $0$ and $1$ for simplicity.
We refer to \citep{Casiraghi2018} for how special cases such as $A_{ij} = \Xi_{ij}$ can be handled.
Again, the entries on the diagonal and in the lower triangular part of $\Omega$ are \(0\) to account for the networks having no self-loops and being undirected.

\subsection{Tractability of the Entropy}\label{sec:tractabilityOfTheEntropy}

\paragraph{Multinomial entropy approximation.}
To compute the Shannon entropy of the fitted gHypEG according to Eqs. \eqref{eq:entropy}, \eqref{eq:walleniusNet}, \eqref{eq:walleniusSOmega} is not straight-forward because of the very large number of networks in this ensemble.
If we were to simply plug the probabilities of all networks into Eq. \eqref{eq:entropy},
the very large number of summands quickly becomes infeasible.
Thus, instead of literally computing the entropy for a fitted gHypEG\@, we compute \(H\) using the fact that, for large networks, the multinomial distribution approximates the Wallenius distribution.
The details of the derivation can be found in App. \ref{sec:convergence}.
Hence,  the gHypEG entropy can be approximated as
\begin{equation}\label{eq:multinomialEntropyApprox}
	H^{\mathrm{mult}} = -\log(m!) - m\sum_{i,j\in V,i<j} p_{ij} \log(p_{ij}) + \sum_{x=2}^m\sum_{i,j\in V,i<j} {m\choose x} p_{ij}^x (1-p_{ij})^{m-x}\log(x!)
\end{equation}
where \(m\) is the number of edges in the network, \(V\) is the set of nodes, and
\begin{equation}\label{eq:multinomialProbabilities}
	p_{ij} = \frac{\Xi_{ij}\Omega_{ij}}{\sum_{kl}\Xi_{kl}\Omega_{kl}}
\end{equation}

\paragraph{Computing the multinomial entropy.}

Equation \eqref{eq:multinomialEntropyApprox} can be computed efficiently even for large ensembles.
In SciPy~\citep{SciPy} there exists an efficient implementation for computing the entropy of a given multinomial distribution.
Our contribution is to apply this to approximate the entropy for a given gHypEG defined by
Eqs. \eqref{eq:multinomialEntropyApprox}, \eqref{eq:multinomialProbabilities}.

\subsection{Comparing Entropy Values}\label{sec:comparingEntropyValues}

\paragraph{Normalizing value ranges.}
The value range of Eq. \eqref{eq:multinomialEntropyApprox} depends on the number of nodes \(n\) and edges \(m\).
In particular, it is a known fact that Shannon entropy attains its maximum value \(H^{\mathrm{max}}\) at equiprobability~\citep{Shannon1948a}.
Hence, the entropy values are always in the interval \([0, H^{\mathrm{max}}]\).

For undirected networks without self-loops equiprobability corresponds to
\begin{equation}\label{eq:pijMaxMultinomialEntropy}
	p_{ij}^{\mathrm{max}} = \frac{2}{n(n-1)}
\end{equation}
i.e., all possible pairs of nodes can be chosen with the same probability.
For two different ensembles, however, $H^{\mathrm{max}}$ can be different because it depends on $n$ and on $m$ via Eq.~\eqref{eq:multinomialEntropyApprox}.
To compare the values of $H^{\mathrm{mult}}$, Eq.~\eqref{eq:multinomialEntropyApprox},  we normalize them by their respective maximum values:
\begin{equation}\label{eq:normalizedEntropy}
	\potentiality{} := \frac{H^{\mathrm{mult}}}{H^{\mathrm{max}}} \equiv \hat{H} \in [0, 1]
\end{equation}
A small value means that the ensemble contains only very few networks, given  the constraints.
With respect to the $p_{ij}$ this means that only very few have probabilities considerably different from zero.
A large value, on the other hand, means that pairs of nodes are chosen almost at random, because of the very few constraints.
Hence, $\hat{H}$ indeed reflects the potentiality of the social organization, namely its ability to attain \emph{different configurations} under given  constraints.

\subsection{Examples for \(\hat{H}\)}\label{sec:syntheticExamples}

\begin{figure}[htbp]
	\centering
	\hfill{}
	\begin{subfigure}[c]{0.3\textwidth}
		\begin{center}
			\resizebox{\textwidth}{!}{

\newcommand{\arrowbundle}[3]{
	\def\increment{15}  
	\pgfmathtruncatemacro\iseven{{Mod(#3,2)}}  
	\ifnum\iseven=0
		\pgfmathsetmacro\baseangle{(\increment / 2) + (((#3 / 2) - 1) * \increment)}
	\else
		\pgfmathsetmacro\baseangle{(#3 - 1) / 2 * \increment}
	\fi

	\pgfmathanglebetweenpoints{\pgfpointanchor{#1}{center}}{\pgfpointanchor{#2}{center}}
	\pgfmathsetmacro\abangle{ \pgfmathresult }  
	\pgfmathanglebetweenpoints{\pgfpointanchor{#2}{center}}{\pgfpointanchor{#1}{center}}
	\pgfmathsetmacro\baangle{ \pgfmathresult }  

	\pgfmathtruncatemacro\lastarrowindex{#3 - 1}
	\foreach \x in {0,...,\lastarrowindex} {
		\pgfmathsetmacro\currangleab{ \abangle + \baseangle - \x * \increment }
		\pgfmathsetmacro\currangleba{ \baangle + \baseangle - (\lastarrowindex - \x) * \increment }
		\draw[arrow] (#1.\currangleab) -- (#2.\currangleba);
	}
}


\begin{tikzpicture}
	\tikzstyle{vertex}=[circle, draw, inner sep=5, fill=orange!70]
	\tikzstyle{arrow}=[thin]

	\pgfmathtruncatemacro\peripheryangle{360 / 10}
	\foreach \n in {1,...,10} {
		\pgfmathtruncatemacro\currangle{\peripheryangle * \n}
		\node[vertex] (n\n) at ($(\currangle:2cm)$) {};
	}

    \foreach \from in {1,...,10} {
        \pgfmathtruncatemacro\linknode{\from + 1}
        \ifnum\linknode<11
            \foreach \to in {\linknode,...,10} {
                \arrowbundle{n\from}{n\to}{2}
            }
        \fi
    }
\end{tikzpicture}
			}
		\end{center}
	\end{subfigure}
	\hfill{}
	\begin{subfigure}[c]{0.3\textwidth}
		\begin{center}
			\resizebox{\textwidth}{!}{

\newcommand{\arrowbundle}[3]{
	\def\increment{10}  
	\pgfmathtruncatemacro\iseven{{Mod(#3,2)}}  
	\ifnum\iseven=0
		\pgfmathsetmacro\baseangle{(\increment / 2) + (((#3 / 2) - 1) * \increment)}
	\else
		\pgfmathsetmacro\baseangle{(#3 - 1) / 2 * \increment}
	\fi

	\pgfmathanglebetweenpoints{\pgfpointanchor{#1}{center}}{\pgfpointanchor{#2}{center}}
	\pgfmathsetmacro\abangle{ \pgfmathresult }  
	\pgfmathanglebetweenpoints{\pgfpointanchor{#2}{center}}{\pgfpointanchor{#1}{center}}
	\pgfmathsetmacro\baangle{ \pgfmathresult }  

	\pgfmathtruncatemacro\lastarrowindex{#3 - 1}
	\foreach \x in {0,...,\lastarrowindex} {
		\pgfmathsetmacro\currangleab{ \abangle + \baseangle - \x * \increment }
		\pgfmathsetmacro\currangleba{ \baangle + \baseangle - (\lastarrowindex - \x) * \increment }
		\draw[arrow] (#1.\currangleab) -- (#2.\currangleba);
	}
}


\begin{tikzpicture}
	\tikzstyle{vertex}=[circle, draw, inner sep=5, fill=orange!70]
	\tikzstyle{arrow}=[thin]

	\pgfmathtruncatemacro\peripheryangle{360 / 9}
	\node[vertex] (center) at (0, 0) {};
	\foreach \n in {2,...,10} {
		\pgfmathtruncatemacro\currangle{\peripheryangle * (\n - 2)}
		\node[vertex] (n\n) at ($(center) + (\currangle:2cm)$) {};
	}

	\foreach \n in {2,...,10} {
		\arrowbundle{center}{n\n}{10}
	}
\end{tikzpicture}
			}
		\end{center}
	\end{subfigure}
	\hfill{}

	\caption{
		The two networks from Sect.~\ref{sec:syntheticExamples}.
		Left: Multi-edge complete network.
		Right: Multi-edge star network.
		\label{fig:starAndER}
	}
\end{figure}
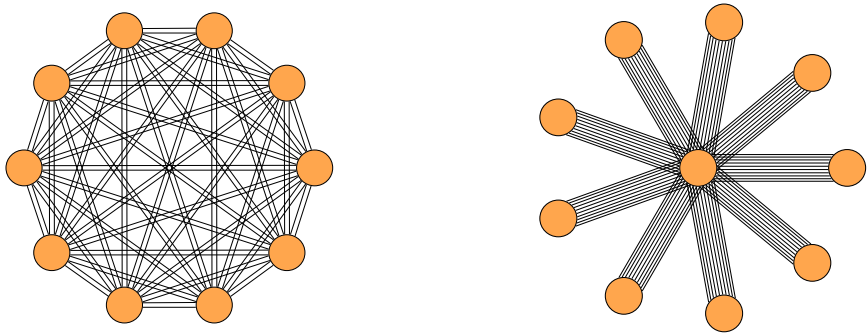

\paragraph{Two special cases.}
To illustrate how constraints can be encoded in the ensemble, we use two examples, a complete network and a star network (see Fig. \ref{fig:starAndER}) for which we consider undirected edges and no self-loops.
We fit the $\Xi$ and $\Omega$ matrices according to Eqs. \eqref{eq:XiComputation}, \eqref{eq:OmegaComputation}.

\paragraph{Complete network.}
This network has \(10\) nodes and \(90\) edges.
Because we consider a multi-edge network, each node has \(2\) edges to every other node.
This results in $d_{i}=d_{j}=18$.
The correction factors are obtained by solving the set of equations in the Appendix~\ref{sec:fittingXi} and yield  $\theta_{i}=\theta_{j}=1.054$.
The resulting network is depicted in Fig. \ref{fig:starAndER}~(left), and the resulting \(\Xi{}\) and \(\Omega{}\) matrices are stated in full in App. \ref{sec:full-matr-compl}.
The entries of the \(\Xi\) matrix for this network are computed according to Eq. \eqref{eq:XiComputation} as
\begin{equation}
	\Xi^{\mathrm{C}}_{ij} =
		\begin{cases}
			720 & \text{if } i < j \\
			0 &	\text{else}
		\end{cases}
\end{equation}
Since in the complete network every node has the same number of edges to every other node, there are no
preferences for specific pairs of nodes.
Therefore one way to choose the \(\Omega\) matrix to encode \emph{no bias} is
\begin{equation}\label{eq:omegaComplete}
	\Omega^{\mathrm{C}}_{ij} =
		\begin{cases}
			1 & \text{if } i < j \\
			0 & \text{else}
		\end{cases}
\end{equation}
which corresponds to Theorem~8 in~\citep{Casiraghi2018} for an undirected network without self-loops.
Remember that \(\Omega_{ij} / \Omega_{kl}\) is the odds-ratio of observing an edge between nodes \(i\) and \(j\) instead of nodes \(k\) and \(l\).
By choosing $\Omega_{ij}$
according to Eq. \eqref{eq:omegaComplete} such ratios are always equal to \(1\).
By plugging \(\Xi^{\mathrm{C}}\) and \(\Omega^{\mathrm{C}}\) into Eq. \eqref{eq:normalizedEntropy} we obtain $\hat{H}=1$.
This means that there are no edge-preferences between particular pairs of nodes, which is trivial because the example was chosen as such.

\paragraph{Star network.}
This network has again \(10\) nodes and \(90\) edges.
But this time there is \(1\) center node and \(9\) peripheral nodes, i.e.,
the network has the constraint that each peripheral node has \(10\) edges that are all attached to the center node, as  depicted in Fig. \ref{fig:starAndER}~(right).
This results in a degree \({d_i=90}\) for the center node placed at \(i=1\) and degrees \({d_j=10}\) for all peripheral nodes \(j\neq 1\).
Again, the \(\Xi\) and \(\Omega\) matrices for this network are stated in full in App. \ref{sec:full-matr-compl}.

When computing the \(\Xi\) matrix according to Eq. \eqref{eq:XiComputation} we obtain
\begin{equation}
	\Xi^{\mathrm{S}}_{ij} =
		\begin{cases}
			3592 & \text{if } i = 1, j > 1 \\
			2 & \text{if } i > 1, j > i \\
			0 &	\text{else}
		\end{cases}
\end{equation}
where the center node is placed at \(i=1\).

The other matrix, \(\Omega\), has to reproduce the constraint that peripheral nodes can only communicate with the center node.
One choice to fulfil this is
\begin{equation}
	\Omega^{\mathrm{S}}_{ij} =
		\begin{cases}
			1 & \text{if } i = 1 \text{ and } i < j \\
			0 &	\text{else}
		\end{cases}
\end{equation}
This choice of \(\Omega\) specifies that observing an edge from node~\(1\) (the center) to any two peripheral nodes \(k\) or \(l\) occurs with the same probability, because the odds-ratio \(\Omega_{1k}/\Omega_{1l}\) is equal to \(1\).
On the other hand, the odds-ratio for a peripheral node \(i\) to form a link with another peripheral node \(k\) instead of with the center node \(1\), namely \(\Omega_{ik}/\Omega_{1i}\), is \(0\) (or infinity if the inverse ratio is formed).
This encodes the constraint that all edges have to be incident to the center node.

By plugging \(\Xi^{\mathrm{S}}\) and \(\Omega^{\mathrm{S}}\) into Eq. \eqref{eq:normalizedEntropy} we obtain $\hat{H}=0.27$ for our star~network.
This value is much lower than for the complete network and reflects the very restrictive constraint that all edges have to be incident to the center node.

\section{Applications to Real-World Data Sets}\label{sec:applications}

\subsection{Examined Data Sets}\label{sec:datasets}

In this Section we apply our potentiality measure to \(5\) empirical networks of social organizations.
These networks were constructed from publicly available data sets which we shortly describe in the following.

\paragraph{Southern Women data set.}
It was introduced by~\citet{davis1941deep} and contains information about \(18\)~women and their participation in \(14\)~social events.
Instead of constructing a bipartite network,
we use a so-called one-mode representation (i.e. a specific projection of the bipartite network) in which the women correspond to the nodes and the edges correspond to co-participations in the social events.
There are no self-loops in this network and edges are undirected.

\paragraph{Karate Club data set.}
It was introduced by~\citet{Zachary1977} and
the network contains \(34\) nodes corresponding to the members of this university Karate club.
Edges correspond to co-participation of members in different activities.
They  are all undirected and there are no self-loops.
There are \(8\) activities considered, thus the number of possible edges between any pair of nodes is less or equal than \(8\).
In total, there are \(231\)~edges.

\paragraph{Conference data set.}
This data set is part of the SocioPatterns project.
It contains data about interactions among conference participants during the ACM~Hypertext~2009 conference.~\citep{Isella2011} To measure the interactions the participants were wearing proximity sensors.
For each interaction between two participants the measured information contains their anonymous ids as well as the time of the respective measurement.
From this information we constructed \(3\) networks, one for each day of the conference.
In each network the nodes correspond to the \(113\) participants in the data and the edges correspond to their interactions at the respective day.
None of the networks contains self-loops and all edges are undirected.
All \(3\) networks have the same set of nodes but differ slightly in the number of edges as can be seen in Table~\ref{table:networkOverview}.

\begin{figure}[htbp]
	\centering
	\hfill{}
	\begin{subfigure}[b]{0.25\textwidth}
		\includegraphics[width=\textwidth]{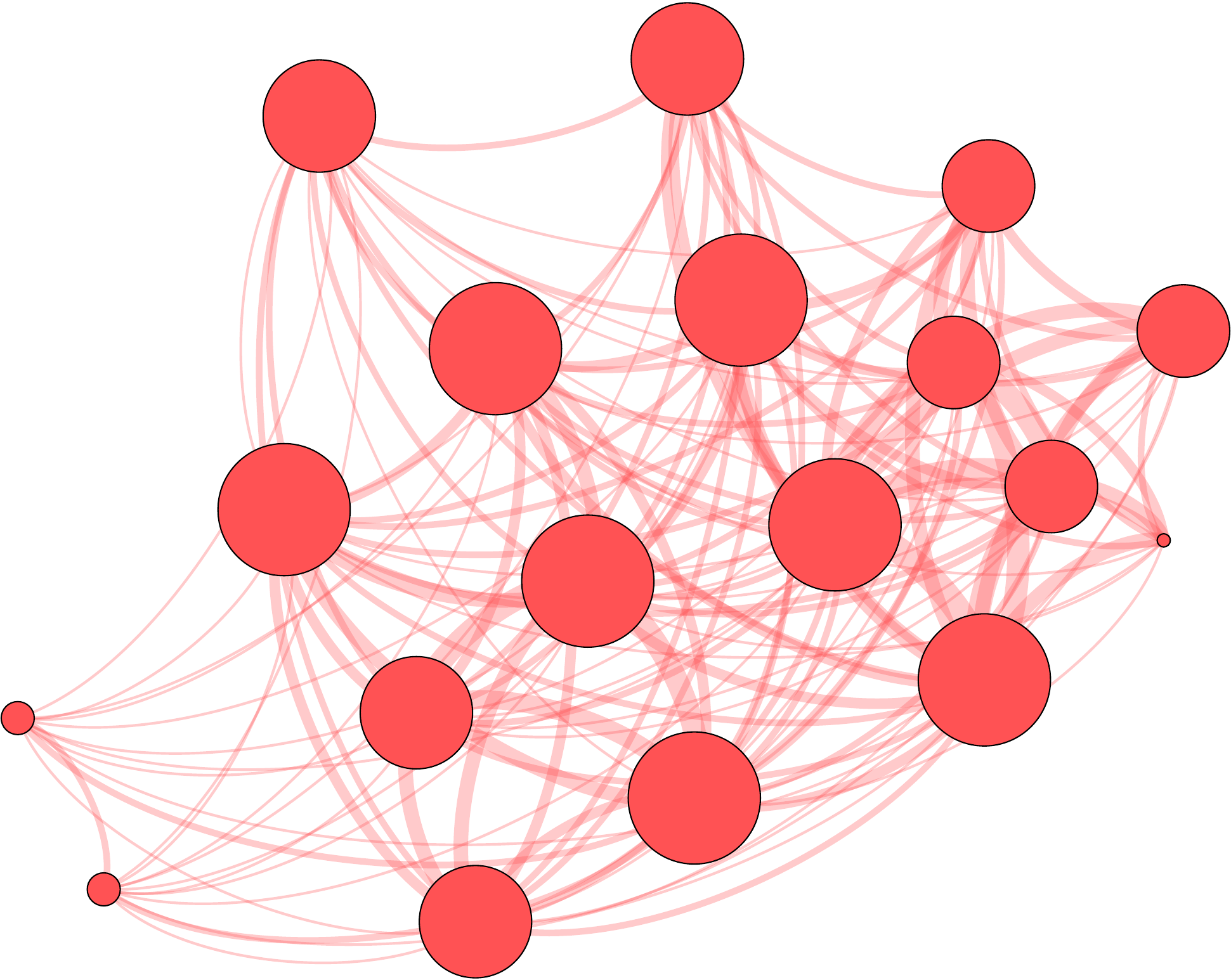}
		\caption{\label{fig:empiricalNetworksSouthernWomen}}
	\end{subfigure}
	\hfill{}
	\begin{subfigure}[b]{0.25\textwidth}
		\includegraphics[width=\textwidth]{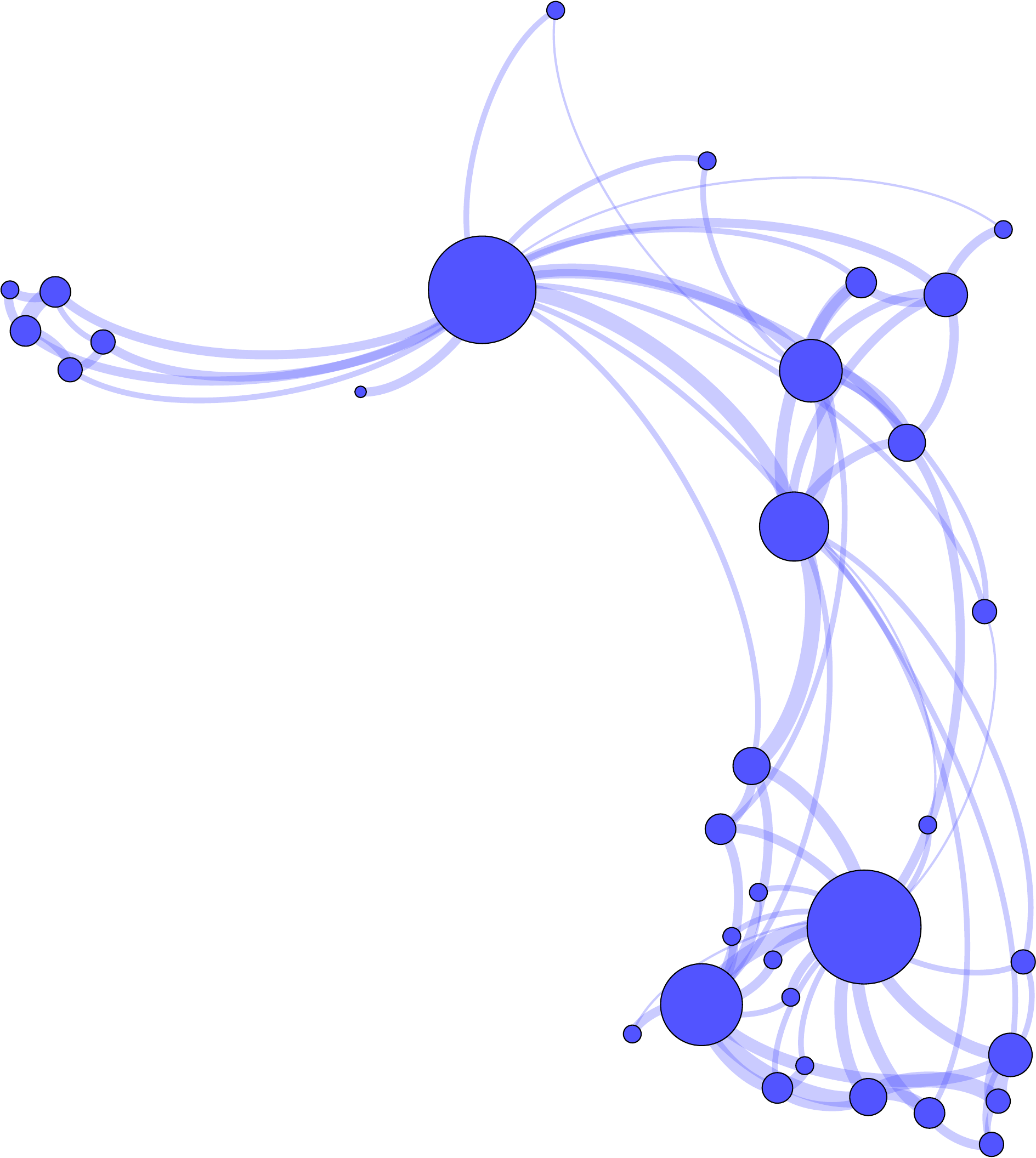}
		\caption{\label{fig:empiricalNetworksKarateClub}}
	\end{subfigure}
	\hfill{}
	\mbox{}
	\\
	\vspace*{1em}
	\begin{subfigure}[b]{0.99\textwidth}
		\centering
		\includegraphics[width=0.25\textwidth]{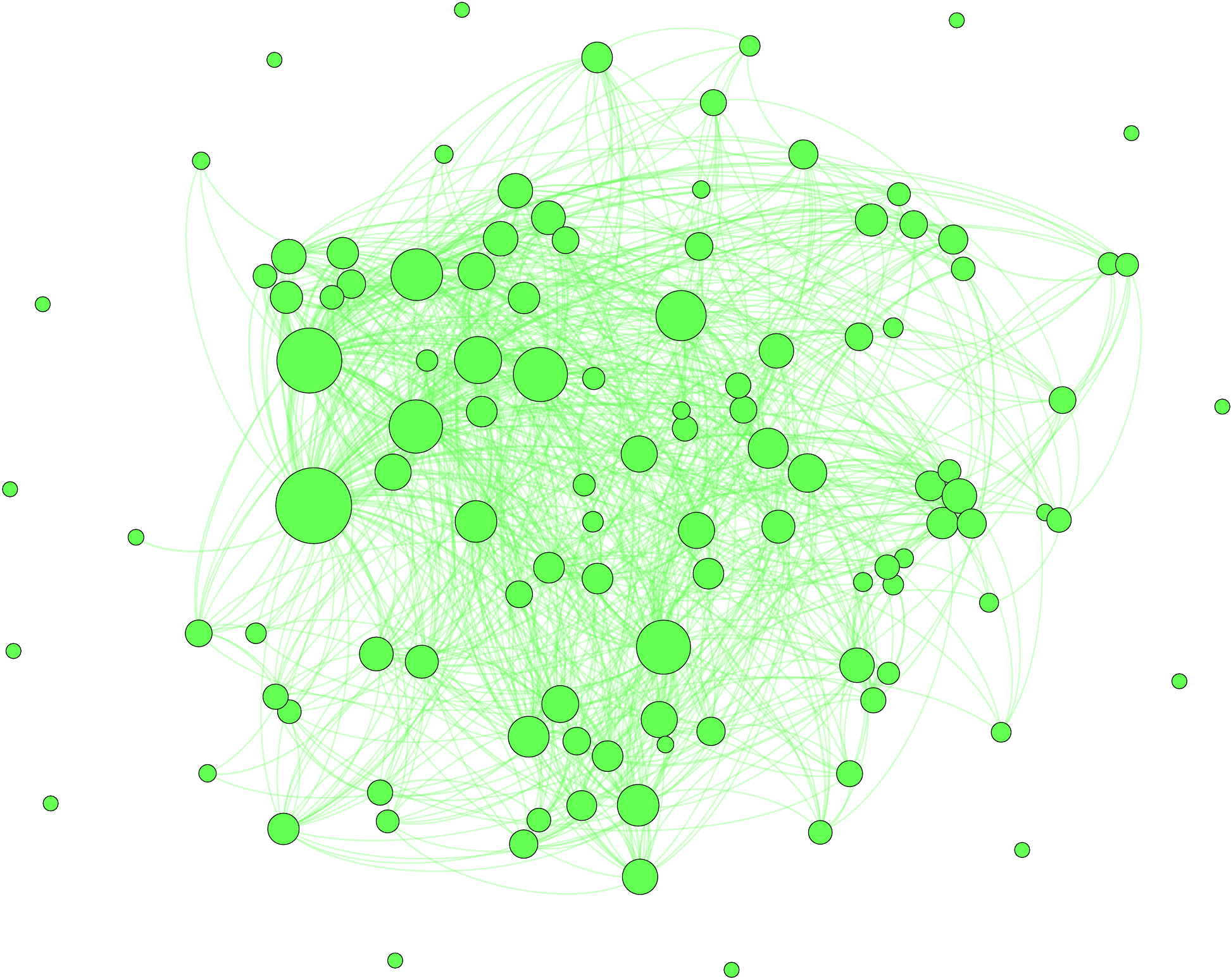}
		\hspace{0.7cm}
		\includegraphics[width=0.25\textwidth]{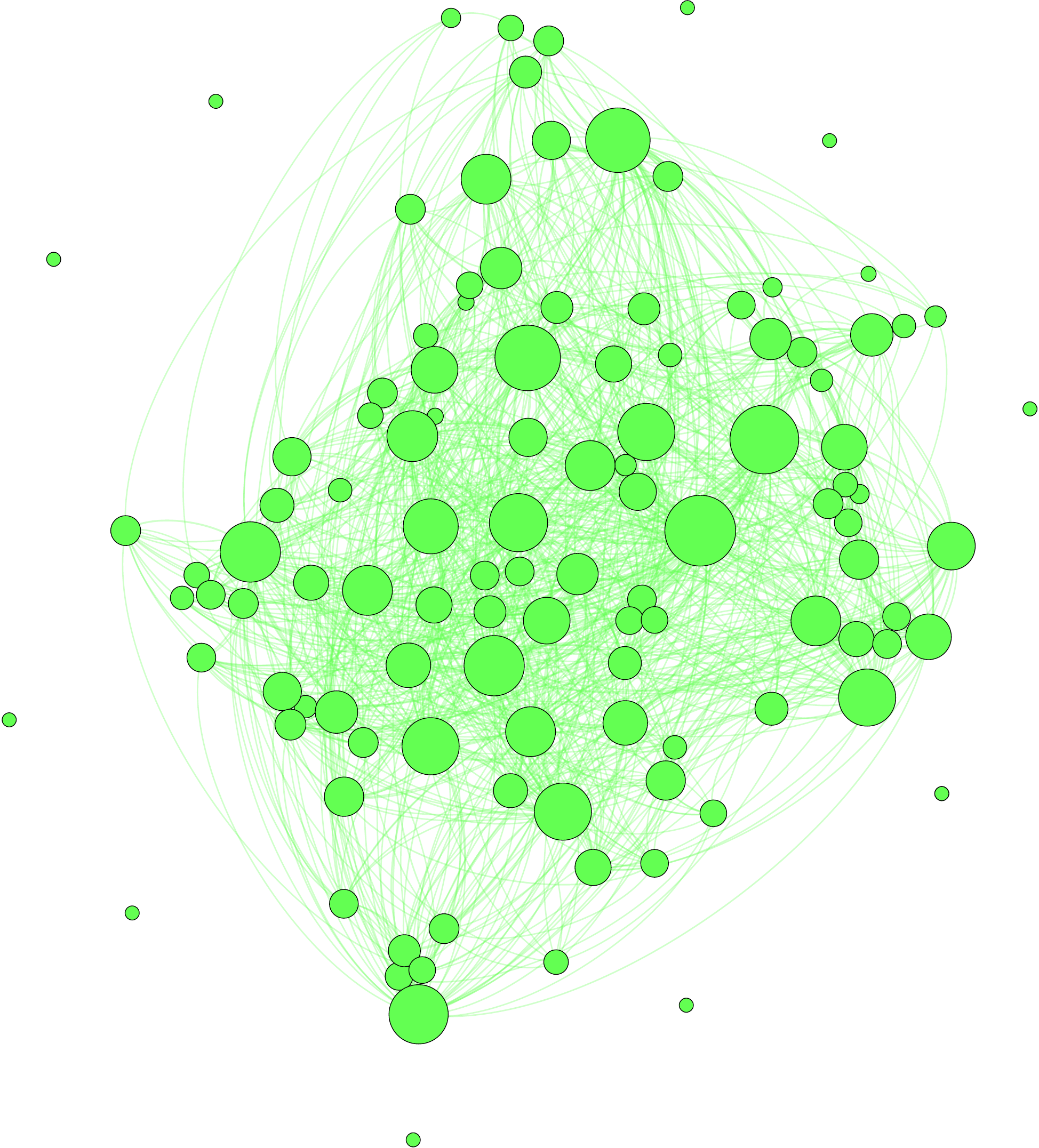}
		\hspace{0.7cm}
		\includegraphics[width=0.25\textwidth]{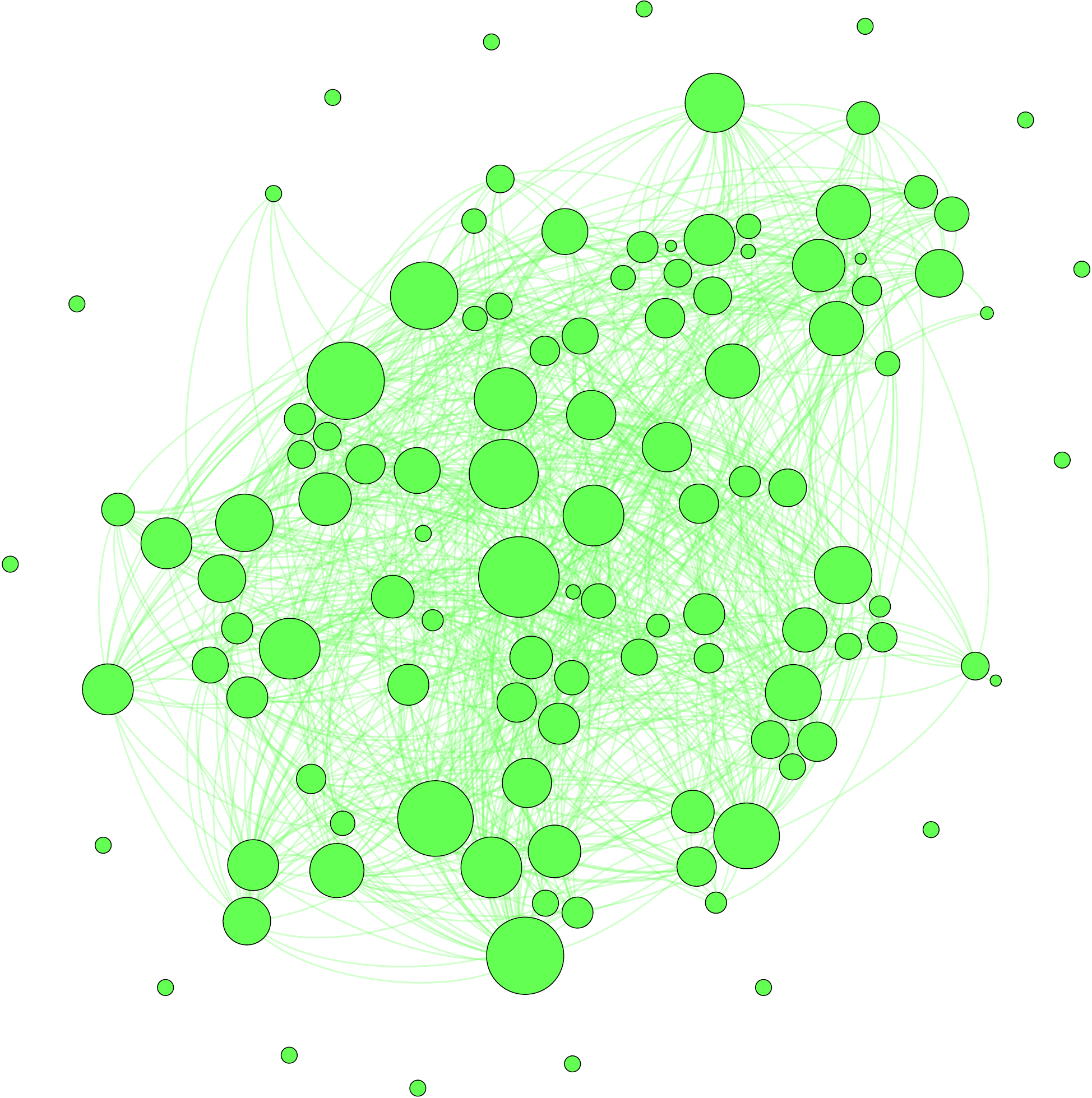}
		\caption{\label{fig:empiricalNetworksHypertextT2}}
	\end{subfigure}
	
	\caption{Network visualizations: \subref{fig:empiricalNetworksSouthernWomen} Southern Women (red),
	\subref{fig:empiricalNetworksKarateClub} Karate Club (blue),
	\subref{fig:empiricalNetworksHypertextT2} Conference (green) at day \(1\) (left), day \(2\) (center), and day \(3\) (right).
		The node size is proportional to the degree.
	}\label{fig:empiricalNetworks}
\end{figure}

\paragraph{Network overview.}
To summarize the networks Table~\ref{table:networkOverview} lists the general network statistics besides the computed potentiality values of $\hat{H}$.
Furthermore, all networks are visualized in Fig. \ref{fig:empiricalNetworks}.
This Figure already suggests that the networks are structurally different.
For example, the Karate Club network shows a cluster structure which is not apparent in the Southern Women network.
And all Conference networks have isolated nodes which neither the Karate Club network nor the Southern Women network have.
\begin{table}[htbp]
	\begin{center}
		{\small
			\begin{tabular}{l|rrrr|rr}
				Network & $n$ & $m$ & $m/n$ & $D$ & $\hat{H}$ & $\hat{H}_{\mathrm{gcc}}$ \\
				\hline
				Southern Women &  $18$ & $322$ &  $17.89$ & $0.91$ & $\mathbf{0.89}$ & $\mathbf{0.89}$ \\
				Karate Club &  $34$ & $231$ &   $6.79$ & $0.14$ & $\mathbf{0.31}$ & $\mathbf{0.31}$ \\
				Conference $t = 1$ & $113$ & $6925$ &  $61.28$ & $0.15$ & $\mathbf{0.21}$ & $\mathbf{0.24}$ \\
				Conference $t = 2$ & $113$ & $7131$ &  $63.11$ & $0.17$ & $\mathbf{0.22}$ & $\mathbf{0.25}$ \\
				Conference $t = 3$ & $113$ & $6762$ &  $59.84$ & $0.15$ & $\mathbf{0.19}$ & $\mathbf{0.23}$ \\
			\end{tabular}
		}
	\end{center}
	\caption{Network statistics of the \(5\) examined empirical networks.
		\(n\) and \(m\) denote the number of nodes and edges in each network.
		\(m/n\) is the average number of multi-edges per node.
		\(D\) is the density of the network, i.e. the number of linked node pairs normalized to the total number of possible node pairs,  after reducing all multi-edges into single edges.
		\(\hat{H}\) denotes the normalized entropy computed according to Eq. \eqref{eq:normalizedEntropy}.
		\(\hat{H}_{\mathrm{gcc}}\) corresponds to \(\hat{H}\) when only the largest connected component in each network is considered.
		All networks are undirected and have no self-loops.
		\label{table:networkOverview}}
\end{table}

\subsection{Potentiality of the Empirical Networks}

For each of the \(5\) empirical networks we computed the potentiality \(\hat{H}\) as outlined in Sects. \ref{sec:potentiality} and \ref{sec:ensembles}.
The computed values for \(\hat{H}\) are listed in Table~\ref{table:networkOverview}.
In the following we comment on the results.

\paragraph{Southern Women network.}
This network attains a very high potentiality at around \(\hat{H}=0.9\), meaning that there are only few constraints in the women's interaction.
In fact, there are almost no preferences for specific pairs of women.
Instead, everyone interacts with everyone else in a rather homogeneous way.
The absence of a preference structure in co-attending events is also visible in the network plot in Fig. \ref{fig:empiricalNetworks}, which looks similar to the complete network considered in Sect. \ref{sec:syntheticExamples}.
This corresponds to the high density, $D$, in Table~\ref{table:networkOverview}.

\paragraph{Karate Club network.}
This network results in a lower potentiality at around \(\hat{H}=0.3\) which indicates that the network is more restricted by constraints.
Indeed, it is known that two social groups, which both had their own leader, co-existed in the Karate club.
Most interactions among the club members occurred within the groups, and in particular with the respective group leaders.
These restrictions explain why the potentiality of this network is not particularly large, especially when compared to the Southern Women network.

\paragraph{Conference networks.}
The lowest potentialities are attained at around \(0.2\) by the  networks of the conference participants.
For each network we observed that the nodes had high degrees, because of the multi-edges, but were linked only to a \emph{few} other nodes (i.e. a rather sparse network).
This implies a relatively strong preferential linkage between \textit{specific} pairs of nodes.
On the other hand, on all \(3\)~days of the conference, there was at least one individual who communicated with at least \(50\%\) of conference participants in the data set (probably the conference organizer).
This induces a star-like interaction effect.
But, there were also a smaller number of isolated nodes which were, on a given day, not involved in any interaction.

Isolated nodes decrease the potentiality because in the theoretical maximum entropy~\(H^{\mathrm{max}}\) they have to be considered. When omitting these isolated nodes, we still find only slightly higher values of the potentialities around \(0.24\)
because of their small number.
Furthermore, it is remarkable that the networks of all three days have similar potentialities.
One could have expected that on day 3 of the conference, participants mainly interact with those they already know.
But this is obviously not the case.

\section{Conclusion}\label{sec:conclusion}

In this paper we address the question how many different states a social organization can attain.
Why is this of importance?
We argue that the number of such possible states is an indication of the ability of the organization to respond to various influences.
As there can be a vast variety of such influences, the corresponding number ideally should be very large.
This indicates that, even for unforeseeable events, the social organization still has many ways to respond.
We call such an ability the \emph{potentiality} of the organization.

To quantify this potentiality, we need an appropriate representation of the social organization.
In this paper, we choose a network approach, where nodes represent individuals and edges their repeated interactions.
This leads to a multi-edge network.
A network ensemble then contains all possible networks that fulfill a given set of constraints.
Such constraints are detected from the observed network and encoded as \emph{propensities}, i.e. as interaction preferences.
The statistical ensemble of all possible networks is then given by the generalized hypergeometric ensemble \mbox{(gHypEG)}.
From this, we can calculate a Shannon entropy, which is used to proxy the potentiality of the organization.

In the following, we comment further on the strengths and weaknesses of our approach.

\paragraph{Fixed numbers of nodes and edges.}

We focus on ensembles with a fixed number of nodes and edges, hence imposing that only networks of this size are attainable by the organization.
Thereby we neglect system growth on purpose, to provide a general measure of potentiality.

\paragraph{Large number of degrees of freedom.}
Using gHypEG, we are able to consider the maximum possible degrees of freedom, meaning that every detail is modeled in the \(\Xi{}\) and \(\Omega{}\) matrices.
This way, we obtain a high model complexity.
A more refined approach could be to compare ensembles of various complexities based on goodness-of-fit measures such as AIC or BIC.
Thereby also simpler ensembles could be involved that, for example, consider communication preferences only between certain \emph{communities} in the network.
Such choices of simpler ensembles were not considered because, again, we want to  provide a general approach not restricted to systems with particular community structures.

\paragraph{Computability.}

For social organizations with only \(10\) individuals and \(10\) interactions there are already more than \(2\cdot 10^{10}\) possible network representations in the ensemble.
According to Eq.~\eqref{eq:entropy}, all of these networks have to be considered individually to compute the Shannon~entropy.
Hence, even simple approaches to directly compute the entropy are  computationally infeasible already for very small organizations.
Our approach instead uses that the Wallenius distribution underlying the gHypEG converges to a multinomial distribution in the limit of large networks for which the entropy can be computed efficiently.
This allows to study the potentiality of a wide range of social organizations.

Our main methodological contribution is indeed the novel way to conceptualize potentiality for a social organization using its representation as a multi-edge network.
As long as this representation is justified, our approach can be extended to other systems.

\bibliographystyle{sg-bibstyle-nourl}
\setlength{\itemsep}{0pt}
\small
\bibliography{bibliography.bib}

\begin{thebibliography}{30}
\expandafter\ifx\csname natexlab\endcsname\relax\def\natexlab#1{#1}\fi
\expandafter\ifx\csname url\endcsname\relax
  \def\url#1{\texttt{#1}}\fi
\expandafter\ifx\csname urlprefix\endcsname\relax\def\urlprefix{URL }\fi
\expandafter\ifx\csname selectlanguage\endcsname\relax
  \def\selectlanguage#1{\relax}\fi

\bibitem[{Bianconi(2008)}]{Bianconi2008}
Bianconi, G. (2008).
\newblock {The entropy of randomized network ensembles}.
\newblock \emph{Europhysics Letters} \textbf{81(2)}, 28005.

\bibitem[{Bianconi(2009)}]{Bianconi2009}
Bianconi, G. (2009).
\newblock {Entropy of network ensembles}.
\newblock \emph{Physical Review E} \textbf{79(3)}, 036114.

\bibitem[{Bollob{\'{a}}s(1998)}]{Bollobas1998}
Bollob{\'{a}}s, B. (1998).
\newblock \emph{{Modern Graph Theory}}, vol. 184 of \emph{Graduate Texts in
  Mathematics}.
\newblock New York, NY: Springer New York.

\bibitem[{Casiraghi and Nanumyan(2018)}]{Casiraghi2018}
Casiraghi, G.; Nanumyan, V. (2018).
\newblock Generalised hypergeometric ensembles of random graphs: the
  configuration model as an urn problem.
\newblock \emph{arXiv preprint arXiv:1810.06495} .

\bibitem[{Casiraghi \emph{et~al.}(2016)Casiraghi, Nanumyan, Scholtes and
  Schweitzer}]{Casiraghi2016}
Casiraghi, G.; Nanumyan, V.; Scholtes, I.; Schweitzer, F. (2016).
\newblock {Generalized Hypergeometric Ensembles: Statistical Hypothesis Testing
  in Complex Networks}.
\newblock \emph{arXiv preprint arXiv:1607.02441} .

\bibitem[{Coon \emph{et~al.}(2018)Coon, Dettmann and Georgiou}]{Coon2018}
Coon, J.~P.; Dettmann, C.~P.; Georgiou, O. (2018).
\newblock {Entropy of spatial network ensembles}.
\newblock \emph{Physical Review E} \textbf{97(4)}, 042319.

\bibitem[{Davis \emph{et~al.}(1941)Davis, Gardner and Gardner}]{davis1941deep}
Davis, A.; Gardner, B.~B.; Gardner, M.~R. (1941).
\newblock \emph{Deep South; a Social Anthropological Study of Caste and Class}.
\newblock Chicago: The University of Chicago Press.

\bibitem[{Dorogovtsev and Mendes(2003)}]{Dorogovtsev2003}
Dorogovtsev, S.; Mendes, J. (2003).
\newblock \emph{{Evolution of Networks}}, vol.~57.
\newblock Oxford University Press.

\bibitem[{Ebeling \emph{et~al.}(1998)Ebeling, Freund and
  Schweitzer}]{Ebeling1998}
Ebeling, W.; Freund, J.; Schweitzer, F. (1998).
\newblock \emph{{Komplexe Strukturen: Entropie und Information}}.
\newblock Wiesbaden: Vieweg+Teubner Verlag.

\bibitem[{Hanel and Thurner(2011)}]{Hanel2011}
Hanel, R.; Thurner, S. (2011).
\newblock {A comprehensive classification of complex statistical systems and an
  axiomatic derivation of their entropy and distribution functions}.
\newblock \emph{Europhysics Letters} \textbf{93(2)}, 20006.

\bibitem[{Isella \emph{et~al.}(2011)Isella, Stehl{\'{e}}, Barrat, Cattuto,
  Pinton and {Van den Broeck}}]{Isella2011}
Isella, L.; Stehl{\'{e}}, J.; Barrat, A.; Cattuto, C.; Pinton, J.-F.; {Van den
  Broeck}, W. (2011).
\newblock {What's in a crowd? Analysis of face-to-face behavioral networks}.
\newblock \emph{Journal of Theoretical Biology} \textbf{271(1)}, 166--180.

\bibitem[{Jones \emph{et~al.}(2001--)Jones, Oliphant, Peterson
  \emph{et~al.}}]{SciPy}
Jones, E.; Oliphant, T.; Peterson, P.; \emph{et~al.} (2001--).
\newblock {SciPy}: Open source scientific tools for {Python}.
\newblock [Online; accessed 2019-01-28].

\bibitem[{Krivitsky and Butts(2017)}]{Krivitsky2017}
Krivitsky, P.~N.; Butts, C.~T. (2017).
\newblock {Exponential-family Random Graph Models for Rank-order Relational
  Data}.
\newblock \emph{Sociological Methodology} , 008117501769262.

\bibitem[{Kulisiewicz \emph{et~al.}(2018)Kulisiewicz, Kazienko, Szymanski and
  Michalski}]{Kulisiewicz2018}
Kulisiewicz, M.; Kazienko, P.; Szymanski, B.~K.; Michalski, R. (2018).
\newblock {Entropy Measures of Human Communication Dynamics}.
\newblock \emph{Scientific Reports} \textbf{8(1)}, 15697.

\bibitem[{Liben-Nowell \emph{et~al.}(2005)Liben-Nowell, Novak, Kumar, Raghavan
  and Tomkins}]{Liben-Nowell2005}
Liben-Nowell, D.; Novak, J.; Kumar, R.; Raghavan, P.; Tomkins, A. (2005).
\newblock {Geographic routing in social networks}.
\newblock \emph{Proceedings of the National Academy of Sciences}
  \textbf{102(33)}, 11623--11628.

\bibitem[{Molloy and Reed(1995)}]{Molloy1995}
Molloy, M.; Reed, B. (1995).
\newblock {A critical point for random graphs with a given degree sequence}.
\newblock \emph{Random Structures {\&} Algorithms} \textbf{6(2-3)}, 161--180.

\bibitem[{Morris \emph{et~al.}(2008)Morris, Handcock and Hunter}]{Morris2008}
Morris, M.; Handcock, M.~S.; Hunter, D.~R. (2008).
\newblock {Specification of Exponential-Family Random Graph Models: Terms and
  Computational Aspects}.
\newblock \emph{Journal of Statistical Software} \textbf{24(4)}.

\bibitem[{Newman(2003)}]{Newman2003}
Newman, M. E.~J. (2003).
\newblock {The Structure and Function of Complex Networks}.
\newblock \emph{SIAM Review} \textbf{45(2)}, 167--256.

\bibitem[{Park and Newman(2004)}]{Park2004}
Park, J.; Newman, M.~E. (2004).
\newblock {Statistical mechanics of networks}.
\newblock \emph{Physical Review E} \textbf{70(6)}, 13.

\bibitem[{Rajaram and Castellani(2016)}]{Rajaram2016}
Rajaram, R.; Castellani, B. (2016).
\newblock {An entropy based measure for comparing distributions of complexity}.
\newblock \emph{Physica A} \textbf{453}, 35--43.

\bibitem[{Santiago-Paz \emph{et~al.}(2012)Santiago-Paz, Torres-Roman and
  Velarde-Alvarado}]{Santiago-Paz2012}
Santiago-Paz, J.; Torres-Roman, D.; Velarde-Alvarado, P. (2012).
\newblock {Detecting anomalies in network traffic using Entropy and Mahalanobis
  distance}.
\newblock In: \emph{CONIELECOMP 2012, 22nd International Conference on
  Electrical Communications and Computers}. IEEE, pp. 86--91.

\bibitem[{Scholtes(2017)}]{Scholtes2017}
Scholtes, I. (2017).
\newblock {When is a Network a Network?}
\newblock In: \emph{Proceedings of the 23rd ACM SIGKDD International Conference
  on Knowledge Discovery and Data Mining - KDD '17}. New York, New York, USA:
  ACM Press, pp. 1037--1046.

\bibitem[{Scholtes \emph{et~al.}(2016)Scholtes, Mavrodiev and
  Schweitzer}]{Scholtes2016}
Scholtes, I.; Mavrodiev, P.; Schweitzer, F. (2016).
\newblock {From Aristotle to Ringelmann: a large-scale analysis of team
  productivity and coordination in Open Source Software projects}.
\newblock \emph{Empirical Software Engineering} \textbf{21(2)}, 642--683.

\bibitem[{Schweitzer \emph{et~al.}(2014)Schweitzer, Nanumyan, Tessone and
  Xia}]{Schweitzer2014}
Schweitzer, F.; Nanumyan, V.; Tessone, C.~J.; Xia, X. (2014).
\newblock {How do OSS projects change in number and size? A large-scale
  analysis to test a model of project growth}.
\newblock \emph{Advances in Complex Systems} \textbf{17(07n08)}, 1550008.

\bibitem[{Shannon(1948)}]{Shannon1948a}
Shannon, C.~E. (1948).
\newblock {A Mathematical Theory of Communication}.
\newblock \emph{Bell System Technical Journal} \textbf{27(3)}, 379--423.

\bibitem[{Vaccario \emph{et~al.}(2018)Vaccario, Verginer and
  Schweitzer}]{Vaccario2018}
Vaccario, G.; Verginer, L.; Schweitzer, F. (2018).
\newblock {Reproducing scientists' mobility: A data-driven model}.
\newblock \emph{arXiv preprint arXiv:1811.07229} .

\bibitem[{Wallenius(1963)}]{wallenius1963}
Wallenius, K.~T. (1963).
\newblock \emph{{Biased Sampling: the Noncentral Hypergeometric Probability
  Distribution}}.
\newblock Ph.d. thesis, Stanford University.

\bibitem[{Zachary(1977)}]{Zachary1977}
Zachary, W.~W. (1977).
\newblock {An Information Flow Model for Conflict and Fission in Small Groups}.
\newblock \emph{Journal of Anthropological Research} \textbf{33(4)}, 452--473.

\bibitem[{Zanetti \emph{et~al.}(2013)Zanetti, Scholtes, Tessone and
  Schweitzer}]{Zanetti2013}
Zanetti, M.~S.; Scholtes, I.; Tessone, C.~J.; Schweitzer, F. (2013).
\newblock {The rise and fall of a central contributor: Dynamics of social
  organization and performance in the GENTOO community}.
\newblock In: \emph{2013 6th International Workshop on Cooperative and Human
  Aspects of Software Engineering, CHASE 2013 - Proceedings}. IEEE, pp. 49--56.

\bibitem[{Zhao \emph{et~al.}(2011)Zhao, Karsai and Bianconi}]{Zhao2011}
Zhao, K.; Karsai, M.; Bianconi, G. (2011).
\newblock {Entropy of dynamical social networks}.
\newblock \emph{PLoS ONE} \textbf{6(12)}, e28116.

\end{thebibliography}

\normalsize
\begin{appendix}

\section*{Appendices}

\section{Fitting the \(\Xi\) Matrix for Undirected Networks Without Self-Loops}\label{sec:fittingXi}

In the case of undirected networks, the \(\Xi{}\) matrix can be derived according to Definition~4 and Lemma~3 in~\citep{Casiraghi2018}, i.e.\ as \(2 d_i d_j\) for the degrees \(d_i\) and \(d_j\) of nodes \(i\) and \(j\).
If we further disallow self-loops, an additional step is necessary to ensure that the expected degrees in the ensemble still correspond to those of the initial network, and the probability spaces are comparable.
In particular, we need to ensure that the entries of the $\mathbf\Xi$ matrix sum to $m^2$ in the case of directed networks, and to $4m^2$ in the case of undirected networks.

To do so, in the case of undirected networks we define the entries of $\mathbf\Xi$ as follows:
\begin{equation}
	\Xi_{ij}:=2d_id_j\theta_i\theta_j,
\end{equation}
where $\theta_i$ is the correction factor corresponding to node $i$.
To estimate the parameters $\theta_i$ we fix the two constraints just described: (i) degrees have to be preserved in expectation, (ii) entries of $\mathbf\Xi$ sum to $4m^2$.
This gives the following system of equations:
\begin{equation}\label{eq:xidef}
	\begin{cases}
		\sum_{i<j}\Xi_{ij}=4m^2 \\
		\frac{m}{\sum_{l<k}\Xi_{lk}}\sum_{j\neq 1}\Xi_{1j}&=d_1\\
		&\vdots\\
		\frac{m}{\sum_{l<k}\Xi_{lk}}\sum_{j\neq n}\Xi_{nj}&=d_n
	\end{cases}
\end{equation}
where $m\sum_j\Xi_{ij}/\sum_{l<k}\Xi_{lk}$ gives the expected degree of node $i$ according to the configuration model~\citep{Casiraghi2018}.
By inputting eq.\ref{eq:xidef} in the system above, we can simplify it into the following system of $n$ equations in $n$ variables for which we find a numerical solution.
\begin{equation}
	\begin{cases}
		2m\theta_1\sum_{j\neq 1}d_j\theta_j&=0\\
		&\vdots\\
		2m\theta_n\sum_{j\neq n}d_j\theta_j&=0
	\end{cases}
\end{equation}
For the case of the \emph{star network} discussed in Sect. \ref{sec:syntheticExamples}, there is no exact solution to this set of equations.
It is not possible to fix the number of edges to \(90\) while simultaneously fixing the expected degrees over a gHypEG ensemble to the observed degrees in the star network.
Therefore, we introduce an approximation that allows for a small error tolerance between a degree \(d_i\) and its corresponding expected degree \(\tilde{d}_i\) in the ensemble:
\begin{equation}
  \abs{d_i-\tilde{d}_i} \leq 0.5  \end{equation}
This allows us to obtain a solution close enough to fulfill the conditions of the equation system.

\section{Convergence in Distribution of gHypEGs}\label{sec:convergence}

In the following discussion, we show a short theorem that provides the limiting distribution for gHypEGs.
Recall that gHypEGs are described by the sampling \textbf{without replacement} of $m$ edges from an urn containing $M = \sum_{ij} \Xi_{ij} = m^2$ edges.
When \(M\) is large, and some other constraints are met, the gHypEGs sampling process can be approximated by a sampling \textbf{with replacement}.
Here, we provide a rigorous demonstration of this statement.
Note that it is a known result that the the hypergeometric distribution (i.e., an urn sampling without replacement) converges to the multinomial distribution (i.e., an urn sampling with replacement).
However, to our best knowledge, the are no analytic proofs that the Wallenius non-central hypergeometric distribution also converges to the multinomial distribution.
Hence, we now prove that the sampling with competition described
by Wallenius' multivariate non-central hypergeometric distribution
can be approximated by a multinomial distribution with probabilities defined as in Eq. \eqref{eq:multinomialProbabilities}.

\paragraph{Theorem: Convergence of Wallenius' distribution to the multinomial distribution}

Let $X$ be a random variable distributed according to Wallenius' multivariate hypergeometric non-central distribution  with parameters $\mathbf\Xi$, $\mathbf\Omega$, and $m$, given by
	\begin{equation}
		P(X=\mathbf{A})=\left[\prod_{i,j\in V}{\dbinom{\Xi_{ij}}{A_{ij}}}\right]
		\int_{0}^{1}{\prod_{i,j\in V}{\left(1-t^{\frac{\Omega_{ij}}{S_{\mathbf{\Omega}} }}\right)^{A_{ij}}}dt}
	\end{equation}
	with
	\begin{equation}
		S_{\mathbf{\Omega}}= \sum_{l,k\in V} \Omega_{lk}(\Xi_{lk}-A_{lk}).
	\end{equation}
	Let $\Xi_{ij} = n \tilde \Xi_{ij} \,\, \forall i,j\in V$ such that
\begin{equation}\label{eq:pijmultinom}
	\frac{\Omega_{ij}\Xi_{ij}}{\sum_{(l,k)\in V\times V}\Omega_{lk}\Xi_{lk}} = \frac{\Omega_{ij}\tilde \Xi_{ij}}{\sum_{(l,k)\in V\times V}\Omega_{lk}\tilde\Xi_{lk}} = p_{ij} \quad \forall i,j\in V.
	\end{equation}
Then, the Wallenius' multivariate non-central hypergeometric distribution converges to the multinomial distribution with probabilities $p_{ij}$:

\begin{equation}\label{eq:multinomial}
P(X=\mathbf A)\to\frac{m!}{\prod_{(i,j)\in V\times V}{A_{ij}!}}\prod_{(i,j)\in V\times V}{\left(p_{ij}\right)^{A_{ij}}} \quad \text{ as } n\to\infty
\end{equation}
\paragraph{Proof.}\hspace{-0.2cm}From now on, we will write $\prod_{(i,j)\in V\times V}$ as $\prod_{i,j\in V}$ and $\sum_{(i,j)\in V\times V}$ as $\sum_{i,j\in V}$.
With this notation the Wallenius' multivariate non-central hypergeometric distribution is

\begin{align}
	\label{proof:wallenius_multi}
	P(X=\mathbf{A})&=
	\left[\prod_{i,j\in V}{\dbinom{\Xi_{ij}}{A_{ij}}}\right]
	\int_{0}^{1}{\prod_{i,j\in V}{\left(1-t^{\frac{\Omega_{ij}}{S_{\mathbf{\Omega}} }}\right)^{A_{ij}}}dt}
\\
\label{proof:wallenius_multi_two_terms}	&=\frac{m!}{M^m}\left[\prod_{i,j\in V}{\dbinom{\Xi_{ij}}{A_{ij}}}\right]
	\cdot \frac{M^m}{m!} \int_{0}^{1}{\prod_{i,j\in V}{\left(1-t^{\frac{\Omega_{ij}}{S_{\mathbf{\Omega}} }}\right)^{A_{ij}}}dt}
\end{align}

where we obtain Eq. \eqref{proof:wallenius_multi_two_terms} by multiplying and dividing by $M^m$ and $m!$.
The first term of Eq. \eqref{proof:wallenius_multi_two_terms} can be re-written in the following form:

\begin{align}
	\frac{m!}{M^m} \prod_{i,j\in V}{\dbinom{\Xi_{ij}}{A_{ij}}} &=\frac{m!}{M^m} \prod_{i,j\in V}{\frac{\Xi_{ij}!}{A_{ij}!\left(\Xi_{ij} - A_{ij} \right)}}\\
	\label{proof:trick_of_m}&= \frac{m!}{\prod_{i,j\in V}{A_{ij}!}}\prod_{i,j\in V}\frac{\Xi_{ij}!}{ M^{A_{ij}} \left(\Xi_{ij} - A_{ij} \right)}\\\label{proof:binexpRes_multi}
	&= \frac{m!}{\prod_{i,j\in V}{A_{ij}!}}\prod_{i,j\in V}\left(\prod_{k = 1}^{A_{ij}}\frac{\Xi_{ij} - A_{ij} + k}{M}\right)
\end{align}

Note that to obtain Eq. \eqref{proof:trick_of_m} we have written $M^m = \prod_{i,j\in V}M^{A_{ij}}$ which follows from the fact that $\sum_{i,j\in V}A_{_ij} = m$.
Let now $\Xi_{ij} = n\tilde\Xi_{ij}$ and $\tilde M=\sum_{i,j\in V}\tilde\Xi_{ij}$ such that $M=\sum_{i,j\in V}\Xi_{ij}=n\sum_{i,j\in V}\tilde\Xi_{ij}=n\tilde M$ and $\tilde\Xi_{ij}/\tilde M=\Xi_{ij}/M=\tilde p_{ij}$.
We substitute $\tilde\Xi_{ij}$ and $\tilde M$ in Eq. \eqref{proof:binexpRes_multi} and then, we can calculate its limit for $n\to\infty$:
\begin{align}
\lim_{n\to\infty}
\frac{m!}{\prod_{i,j\in V}{A_{ij}!}}\prod_{i,j\in V}
\left(
	\prod_{k = 1}^{A_{ij}}\frac{n\tilde\Xi_{ij} - A_{ij} + k}{n\tilde M}
\right)&=\frac{m!}{\prod_{i,j\in V}{A_{ij}!}}\prod_{i,j\in V}
\left(
	\prod_{k = 1}^{A_{ij}}
		\lim_{n\to\infty}\frac{n\tilde\Xi_{ij} - A_{ij} + k}{n\tilde M}
\right)
\\
&=\frac{m!}{\prod_{i,j\in V}{A_{ij}!}}\prod_{i,j\in V} \tilde p_{ij} ^{A_{ij}}\label{proof:first_term_multi}
\end{align}
We are left with second term of Eq. \eqref{proof:wallenius_multi_two_terms} that contains the integral.
To evaluate this term, we substitute $\Xi_{ij} = n \tilde\Xi_{ij}$ and $M = n\tilde M$ and calculate its limit for $n\to\infty$:
\begin{align}
	\lim_{n\to\infty}
	\frac{M^m}{m!} \int_{0}^{1}{
		\prod_{i,j\in V}{\left(1-t^{\frac{\Omega_{ij}}{S_{\mathbf{\Omega}} }}\right)^{A_{ij}}}dt
		}&=
\lim_{n\to\infty}\frac{\left(n\tilde M\right)^m}{m!}
\int_{0}^{1}{
	\prod_{i,j\in V}{
		\left(
			1-t^{\frac{\Omega_{ij}}{\sum_{lk}{ \Omega_{lk}(n\tilde \Xi_{lk}-A_{lk}) }} }
		\right)^{A_{ij}}
	}
dt}\\
\label{proof:lim2_multi}&=\frac{\tilde M ^m}{m!} \cdot
\int_{0}^{1}{\lim_{n\to\infty}n^m
	\prod_{i,j\in V}{
		\left(
			1-t^{\frac{\Omega_{ij}}{\sum_{lk}{ \Omega_{lk}(n\tilde \Xi_{lk}-A_{lk}) }} }
		\right)^{A_{ij}}
	}
dt}\\
&=
\frac{\tilde M ^m}{m!} \cdot
\int_{0}^{1}{
	\prod_{i,j\in V}{\lim_{n\to\infty}n^{A_{ij}}
		\left(
			1-t^{\frac{\Omega_{ij}}{\sum_{lk}{ \Omega_{lk}(n\tilde \Xi_{lk}-A_{lk}) }} }
		\right)^{A_{ij}}
	}
dt}\\
& = \frac{\tilde M ^m}{m!} \cdot
\int_{0}^{1}{
	\prod_{i,j\in V}{
			\left(
				\lim_{n\to\infty} n\left(1-t^{\frac{\Omega_{ij}}{\sum_{lk}{ \Omega_{lk}(n\tilde \Xi_{lk}-A_{lk}) }} }\right)
			\right)^{A_{ij}}
	}
dt}\\
\label{proof:lim3_multi}& =\frac{\tilde M ^m}{m!} \cdot
\int_{0}^{1}{
	\prod_{i,j\in V}{
		\left(
			-\frac{\Omega_{ij} \log (t)}{\sum_{lk} \tilde\Xi_{lk} \Omega_{lk}}
		\right)^{A_{ij}}
	}
dt}\\
& =\prod_{i,j\in V}{ \left(
	\frac{\Omega_{ij} \tilde M }{\sum_{lk} \tilde\Xi_{lk} \Omega_{lk}}
\right)^{A_{ij}}}
\cdot
\frac{1}{m!} \int_{0}^{1}{\left(\log{\frac{1}{t}}\right)^{m}}\\
& =\prod_{i,j\in V}{ \left(
	\frac{\Omega_{ij} \tilde M }{\sum_{lk} \tilde\Xi_{lk} \Omega_{lk}}
\right)^{A_{ij}}}\label{proof:second_term_multi}
\end{align}
Note that Eq. \eqref{proof:lim2_multi} follows by the Lebesgue dominated convergence theorem and by the finiteness of the factors in the integral.
To obtain Eq. \eqref{proof:lim3_multi} we have used l'H\^opital's rule, i.e. by recalling that $\lim_{n\to \infty} (1-t^{a/n})n = \lim_{x\to 0} (1-t^{ax})/x  = \lim_{x\to 0} (-a\log(t)t^{ax}) =-a\log(t)$.
Equation \eqref{proof:second_term_multi} follows from an integral definition of the $\Gamma$ function, precisely we have used $\Gamma(z+1) = \int_0^1\log(1/t)^{z}dt = z!$.
Finally, by joining Eq. \eqref{proof:first_term_multi} and Eq. \eqref{proof:second_term_multi}, we obtain the limit of Eq. \eqref{proof:wallenius_multi}:
\begin{align}
	\frac{m!}{\prod_{i,j\in V}{A_{ij}!}}\prod_{i,j\in V} \tilde p_{ij} ^{A_{ij}}
	\prod_{i,j\in V}{ \left(
		\frac{\Omega_{ij} \tilde M }{\sum_{lk} \tilde\Xi_{lk} \Omega_{lk}}
	\right)^{A_{ij}}}
&=\frac{m!}{\prod_{i,j\in V}{A_{ij}!}}
\prod_{i,j\in V} \left(
		\tilde p_{ij}
		\cdot
		\frac{\Omega_{ij} \tilde M }{\sum_{lk} \tilde\Xi_{lk} \Omega_{lk}}
	\right)^{A_{ij}}& \\
\label{proof:p_tilde_def_multi} &=\frac{m!}{\prod_{i,j\in V}{A_{ij}!}}
\prod_{i,j\in V} \left(
		\frac{\Omega_{ij} \tilde\Xi_{ij} }{\sum_{lk} \tilde\Xi_{lk} \Omega_{lk}}
	\right)^{A_{ij}}& \\
\label{proof:p_def_multi}&=\frac{m!}{\prod_{i,j\in V}{A_{ij}!}}
	\prod_{i,j\in V} \left(
		p_{ij}
		\right)^{A_{ij}}
\end{align}
where Eq. \eqref{proof:p_tilde_def_multi} follows from the definition of $\tilde{p}_{ij} = \tilde\Xi_{ij} / \tilde{M}$
and Eq. \eqref{proof:p_def_multi} follows from the definition of $p_{ij}$.$\square$

\section{Full Matrices for the Complete and the Star Network}
\label{sec:full-matr-compl}

This appendix contains the full \(\Xi\) and \(\Omega\) matrices of the gHypEG fits for the two networks described in Sect. \ref{sec:syntheticExamples}.

The matrices of the complete network are
\begin{equation}\label{eq:xiOmegaComplete}
	\Xi^{\mathrm{C}} =
		\begin{pmatrix}
			0 &  720 &  720 &  720 &  \dots &  720 \\
			0 &    0 &  720 &  720 &   &   \\
			0 &  0 &    0 &  720 &   &  \vdots \\
			0 &  0 &  0 &    0 &  \ddots &   \\
			\vdots &   &   &  \ddots &  \ddots &  720 \\
			0 &   &  \dots &   &  0 &    0 \\
		\end{pmatrix}
	\qquad
	\quad
	\Omega^{\mathrm{C}} =
		\begin{pmatrix}
			0 &    1 &    1 &    1 &    \dots &    1 \\
			0 &    0 &    1 &    1 &    &     \\
			0 &    0 &    0 &    1 &    &    \vdots \\
			0 &    0 &    0 &    0 &    \ddots &     \\
			\vdots &     &     &  \ddots &    \ddots &    1 \\
			0 &     &    \dots &    &    0 &    0 \\
		\end{pmatrix}
\end{equation}

The matrices of the star network are
\begin{equation}\label{eq:xiOmegaStar}
	\Xi^{\mathrm{S}} =
		\begin{pmatrix}
			0 & 3592 & 3592 & 3592 &  \hdots & 3592 \\
			0 &    0 &    2 &    2 &  \hdots &    2 \\
			0 &    0 &    0 &    2 &   &   2 \\
			0 &    0 &    0 &    0 &  \ddots &  \vdots \\
			\vdots   &      &      &  \ddots &  \ddots &  2 \\
			0 &      &  \hdots &    &  0 &    0 \\
		\end{pmatrix}
	\qquad
	\quad
	\Omega^{\mathrm{S}} =
		\begin{pmatrix}
			0 &    1 &    1 &    1 &    \hdots &    1 \\
			0 &    0 &    0 &    0 &    \hdots &    0 \\
			0 &    0 &    0 &    0 &    &    \\
			0 &    0 &    0 &    0 &    &    \vdots  \\
			\vdots &    &    &    &    \ddots &    0 \\
			0 &    &    \hdots &     &    0 &    0 \\
		\end{pmatrix}
\end{equation}
The \(\Xi^{\mathrm{S}}\) is only an approximation as outlined in App. \ref{sec:fittingXi}.
It is needed because the equation-system to determine the correction factors has no exact solution for the star network.

\end{appendix}

\end{document}